\documentclass[10pt,conference]{IEEEtran}
\IEEEoverridecommandlockouts
\usepackage{cite}
\usepackage{amsmath,amssymb,amsfonts}
\usepackage{algorithmic}
\usepackage{graphicx}
\usepackage{textcomp}
\usepackage{xcolor}
\usepackage{xurl}
\def\BibTeX{{\rm B\kern-.05em{\sc i\kern-.025em b}\kern-.08em
    T\kern-.1667em\lower.7ex\hbox{E}\kern-.125emX}}

\usepackage{array}

\newcolumntype{L}[1]{>{\raggedright\arraybackslash}p{#1}}
\newcolumntype{C}[1]{>{\centering\arraybackslash}p{#1}}
\newcolumntype{R}[1]{>{\raggedleft\arraybackslash}p{#1}}
\newcolumntype{M}[1]{>{\centering\arraybackslash}m{#1}}

\usepackage{multirow}
\usepackage{multicol}
\usepackage[most]{tcolorbox}
\usepackage[normalem]{ulem}

\begin{document}

\title{From Research to Practice: An Interactive Rapid Review of Autonomous Driving System Testing in Industry\\

\thanks{This work was supported in part by the Wallenberg Foundation and WASP Postdoctoral Scholarship Program - KAW 2023.0474.}
}


\author{
\IEEEauthorblockN{
Qunying Song\IEEEauthorrefmark{1},
Ali Nouri\IEEEauthorrefmark{2},
Håkan Sivencrona\IEEEauthorrefmark{2},
Federica Sarro\IEEEauthorrefmark{1}
}
\IEEEauthorblockA{
\IEEEauthorrefmark{1}University College London, London, UK\\
\IEEEauthorrefmark{2}Volvo Cars, Gothenburg, Sweden\\
Email: \{qunying.song, f.sarro\}@ucl.ac.uk,
\{ali.nouri, hakan.sivencrona\}@volvocars.com
}
}

\maketitle

\begin{abstract}
Autonomous driving systems (ADS) are increasingly deployed in real traffic, yet testing remains fundamentally challenging due to open environments, complex scenarios, and the lack of established processes and acceptance criteria. Despite extensive research, a gap persists between academic advances and their applicability in industrial practice. To address this, we conduct an interactive rapid review in collaboration with 21 practitioners from a leading automotive company. Practitioners identified 12 key challenges in ADS testing, and prioritised two as the most critical issues, namely approaches to and completeness of testing for End-to-End (E2E) ADS. We analyzed 17 research studies relevant to these two challenges, most of which focus on generating critical testing scenarios, and subsequently assessed their relevance and applicability in practice. Our study provides the first practitioner-driven review and evaluation of current ADS testing research, reveals practical challenges in ADS testing, offers rapid insights for practitioners, and highlights the need for more context-aware, industry-relevant solutions to bridge the gap between research and practice.
\end{abstract}


\section{Introduction}
\label{sec:intro}

Autonomous driving systems (ADS) have evolved rapidly in recent years and are now deployed in commercial robotaxi services across regions, such as the United States and China~\cite{liao2025advancing}. The deployment and safe operation of such systems require extensive and effective testing to validate their behavior across a wide range of driving scenarios within their operational environments~\cite{riedmaier2020survey, sun2021scenario}. Although research on ADS testing is growing, industrial testing practices, particularly real-world challenges and their specific contexts, are rarely reported in the literature~\cite{song2024empirically}. Moreover, academic research on ADS testing is not always evaluated or validated in industrial settings, and the applicability of research results in practice is often overlooked, ultimately limiting their practical value and impact~\cite{song2025generative}.

To identify pressing testing challenges in the autonomous driving industry, explore relevant research, and assess how proposed interventions support practitioners in practice, we collaborate with an industry partner and conduct an \textit{interactive rapid review}~\cite{rico2020guidelines}. We address the following research questions in the context of our case company:

\newpage
\begin{enumerate}
\item[RQ1] What are the key challenges in testing ADS in practice?
\item[RQ2] Which research is relevant to these challenges?
\item[RQ3] How feasible are these approaches in practice, and what constraints affect their adoption?
\end{enumerate}

We adopt an interactive rapid review~\cite{rico2020guidelines} approach, involving 21 practitioners from the case company. Practitioners first proposed challenges in ADS testing. Based on their prioritization, we focused on the two most important challenges, both related to testing End-to-End (E2E) ADS, and selected relevant research studies through a systematic literature search. After analyzing these studies and extracting technological rules~\cite{runeson2020design} that summarize their interventions, we presented the findings to practitioners to assess applicability and understand factors influencing adoption in their industrial context.

The main contributions of this study are fourfold:

\begin{itemize}
    \item \textbf{Practical challenges.} We identify 12 practical challenges faced by industry practitioners, revealing key problem areas and future directions in ADS testing.
    \item \textbf{Research synthesis.} We select 17 research studies from an initial pool of 1,389 papers and extract technological rules to synthesize their contributions.
    \item \textbf{Knowledge transfer.} We present the technological rules and findings to practitioners to support knowledge transfer and rapid decision-making.
    \item \textbf{Industrial relevance.} We provide insights into their industrial relevance by gathering practitioner feedback on their applicability and factors influencing adoption.
\end{itemize}

Beyond providing rapid insights and potential solutions for testing E2E ADS in industry, this study provides an industrial lens for interpreting academic research and demonstrates the importance of \textit{context-aware approaches} in software engineering~\cite{basili2018software, briand2017case}. These approaches are essential for improving research relevance and facilitating effective technology transfer. To the best of our knowledge, this is the first study to apply an interactive rapid review to ADS testing.


\section{Related Work}
\label{sec:related_work}

Several surveys have examined challenges, practices, approaches, and future trends in ADS testing~\cite{beringhoff2022thirty, song2024empirically, tang2023survey, riedmaier2020survey, liao2025advancing, lou2022testing}, while others have aggregated tools, datasets, simulators, and platforms~\cite{ji2021perspective, kang2019test, ma2021traffic, rosique2019systematic}. These studies provide overviews of the state of the art and resources relevant to both researchers and practitioners. More recently, some studies have focused specifically on the use of generative AI for ADS testing~\cite{song2025generative, zhao2026survey, tian2025large, gao2026foundation, wu2026foundation}. These reviews vary in methodology, with some relying primarily on academic literature~\cite{tang2023survey, riedmaier2020survey, zhang2022finding, cai2022survey, song2025generative, gao2026foundation, wu2026foundation, tian2025large, zhao2026survey}, others on public sources~\cite{ji2021perspective, kang2019test, ma2021traffic, rosique2019systematic}, some on practitioner input (e.g., interviews~\cite{song2024empirically, song2024industry, beringhoff2022thirty}), and some using hybrid approaches~\cite{liao2025advancing, lou2022testing, knauss2017paving}. However, to our knowledge, no existing review has adopted an interactive rapid review approach that involves practitioners throughout the process and systematically evaluates research interventions against current industrial challenges. This study is the first to apply such an approach to ADS testing. Unlike conventional reviews, which are primarily literature-driven, our study is grounded in practitioner-identified challenges and includes post hoc assessments of relevance and applicability in an industrial context. Furthermore, no prior review has specifically focused on testing E2E ADS, which was identified here as one of the most urgent and important challenges by practitioners, further strengthening the novelty and contributions of this work.

\section{Research Method}
\label{sec:method}

\begin{figure*}[tbp]
    \centering
    \includegraphics[width=\textwidth, trim=0 50mm 0 50mm, clip, width=\textwidth]{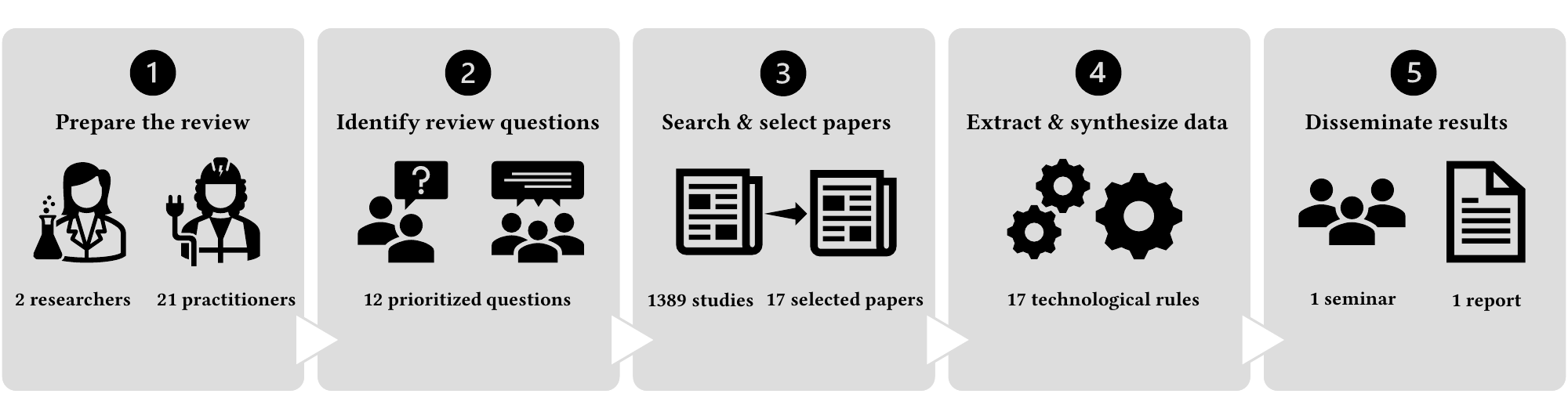}
        \caption{An overview of our research method, including the five key steps along with their main elements and outputs.}
    \label{fig:methodology}
\end{figure*}

We referred to the guidelines proposed by Rico et al.~\cite{rico2020guidelines} for conducting interactive rapid reviews in software engineering, building on our prior experience with this approach~\cite{song2022exploring}. An interactive rapid review aims to quickly synthesize existing knowledge to support decision-making, while ensuring continuous involvement of practitioners throughout the process.

We collaborated with our case company for this review because it is a long-established automotive company with extensive experience in the development and testing of ADS. In addition, the company was readily accessible for collaboration and demonstrated a strong commitment to the study. An overview of the research method is presented in Figure~\ref{fig:methodology}.

\subsection{Preparing the Review}
\label{sec:method:prepare}

The first step of the interactive rapid review is to establish a review team of researchers and practitioners, identify shared interests, and define a collaboration protocol. To achieve this, we organized a workshop with two practitioners from the case company. During the session, the first author presented the study’s objectives, namely to explore interventions for ADS testing reported in the academic literature and to assess their applicability in practice, and introduced the interactive rapid review approach. Following the workshop, the team agreed on a collaboration protocol specifying communication channels, meeting frequency, and expectations for both researchers and practitioners at each stage of the study.

\subsection{Identifying Review Questions}
\label{sec:method:questions}

The second step of the interactive rapid review is to identify the review questions, which guide the search for and selection of relevant literature. To this end, we organized two workshops. In the first workshop, four practitioners proposed challenges and open questions related to ADS testing, describing the specific concerns and contexts associated with each issue. In the second workshop, five practitioners prioritized these challenges through a voting process to identify those considered most urgent and critical. To mitigate potential bias, four of the participants involved in the voting had not contributed to the initial questions. Challenges receiving more votes were ranked higher. As a result, we obtained a prioritized list of 12 review questions, as presented in Table~\ref{tab:questions}, and selected the two highest-ranked questions for this review.

\subsection{Searching for and Selecting Papers}
\label{sec:method:papers}

After identifying the review questions, the next step is to search for and select relevant papers. We began by extracting search keywords from the questions, as shown in Table~\ref{tab:keywords}. We then queried several major research databases, including IEEE Xplore~\cite{ieee}, ACM Digital Library~\cite{acm}, Scopus~\cite{scopus}, and arXiv~\cite{arxiv}. Unlike a systematic literature review, which aims to identify all relevant studies, an interactive rapid review focuses on identifying the most relevant papers that directly address practitioners’ needs and enable timely insights. We evaluated the initial search results based on their titles, abstracts, introductions, and conclusions. When a paper's relevance was unclear, we consulted practitioners to clarify industrial objectives, contextual factors, or constraints, to inform inclusion or exclusion decisions. In total, 17 papers were deemed relevant to the two review questions, as shown in Table~\ref{tab:results}.

\subsection{Extracting and Synthesizing Data}
\label{sec:method:synthesize}

From the selected papers, we extracted 17 technological rules, one per paper. A technological rule, as described by Runeson et al.~\cite{runeson2020design}, represents a research contribution in a structured form, capturing the intervention, context, and expected effect. Such rules can be formulated at different levels of abstraction and serve to communicate research outcomes in a concise and accessible manner. The extracted rules were subsequently presented to 21 practitioners, including 8 who had participated in question formulation and prioritization and 13 newly involved, in a follow-up workshop. During this session, practitioners reflected on their relevance and applicability within the specific context of the case company; the results of this assessment are reported in Section~\ref{sec:results:assessments}.

\subsection{Disseminating Review Results}
\label{sec:method:disseminate}
The final step of the interactive rapid review is disseminating the results. On the practitioner side, we presented the findings, including the technological rules extracted from the selected papers, to stakeholders at a workshop within the case company, as described in Section~\ref{sec:method:synthesize}. The workshop facilitated knowledge transfer and supported continued collaboration. On the academic side, the findings were synthesized and reported in this paper, providing a consolidated account of the technological rules and their assessments.

\section{Review Questions}
\label{sec:results:questions}


\begin{table*}[tbp]
\centering
\caption{Review questions proposed by practitioners from the case company, together with their mean ratings for importance and urgency. The two highest-rated questions selected for this review are highlighted in bold.}
\begin{tabular}{|C{0.02\textwidth}|L{0.85\textwidth}|C{0.05\textwidth}|}
    \hline 
    \textbf{\#}  & \textbf{Review Question} & \textbf{$\bar{rating}$} \\
    \hline
    1 & How to argue on completeness of verification and validation for ADS? & 3.40 \\
    \hline
    2 & How to handle infinite state space of test scenarios? & 3.20 \\
    \hline
    3 & How to systematically define scenario-based testing? & 3.40 \\
    \hline
    4 & What coverage metrics can be used? & 3.20 \\
    \hline
    5 & How can we ensure sufficient scenario coverage for the ODD and the long tail? & 2.80 \\
    \hline
    6 & What quantitative coverage metrics define “enough testing” for closed-loop, multi-agent driving? & 2.80 \\
    \hline
    7 & \textbf{How does architecture affect the verification and validation strategy (e.g., E2E vs Modular approach)?} & \textbf{4.40} \\
    \hline
    8 & \textbf{How to argue on completeness of verification and validation for an E2E-enabled ADS?} & \textbf{4.40} \\
    \hline
    9 & What to adapt conventional verification and validation methods such as unit testing with E2E? & 4.00 \\
    \hline
    10 & How to test the data quality? Data is requirement and hence instead of verification and validation of requirement here we need to test the quality of data during whole DataOps process? & 3.60 \\
    \hline
    11 & How can we build reusable safety evidence across rapid ML iterations and OTA updates? & 3.80 \\
    \hline
    12 & How do we validate simulation (Game engine based, 3DGS, Nerf, Diffusion) credibility and manage the sim-to-real gap? & 4.20 \\
    \hline
\end{tabular}
\label{tab:questions}
\end{table*}

\subsection{Proposed Questions}
\label{sec:results:questions:proposed}

As described in Section~\ref{sec:method:questions}, four practitioners from the case company, representing diverse roles in ADS testing, participated in the initial workshop and proposed a total of 12 review questions (Table~\ref{tab:questions}). For clarity, these questions span three main themes:

\begin{itemize}
    \item \textbf{Testing completeness (Q1--Q6):} questions concerning testing completeness, including demonstrating completeness, defining the scenario space, establishing quantitative metrics, and determining target coverage.
    \item \textbf{E2E ADS testing (Q7--Q9):} questions examining how system architecture influences testing approaches and the evaluation of testing completeness for E2E ADS.
    \item \textbf{Other testing aspects (Q10--Q12)}: questions concerning data quality validation, safety argumentation, and simulation realism. 
\end{itemize}

\subsection{Selected Questions}
\label{sec:results:questions:selected}

After the questions were formulated, five practitioners participated in a second workshop to prioritize them. To mitigate potential bias in the assessment of importance and urgency, a different group of experts was deliberately involved in the rating process, with only one participant having contributed to the formulation of the questions. The questions were rated on a \textit{five-point Likert scale} ranging from “not important” to “very important.” Additionally, an open-ended question (Q13) was included to capture any further issues beyond those initially proposed. The results indicate that Q7 (the impact of transitioning to E2E architecture on ADS testing) and Q8 (approaches to arguing for the completeness of testing for E2E ADS) received the highest mean ratings, both scoring 4.40. More broadly, all three E2E ADS testing questions (Q7–Q9) were rated highly, indicating substantial interest in and demand for research on this architectural paradigm. To focus the review on the most important issues and to provide timely insights to practitioners, we focused on the two highest-rated questions.

\section{Literature Search}
\label{sec:results:papers}


\begin{table*}[tbp]
\centering
\caption{Search keywords derived from the review questions. Bold, solid-underlined text indicates extracted keywords; grey, dashed-underlined text indicates terms removed after the preliminary search. }
\begin{tabular}{|C{0.03\textwidth}|L{0.31\textwidth}|C{0.20\textwidth}|C{0.19\textwidth}|}
    \hline 
    \textbf{\#}  & \textbf{Review Question} & \textbf{Common Keywords} & \textbf{Specific Keywords}\\
    \hline
    \multirow{3}{*}{Q7} & How does architecture \textcolor{gray}{\dashuline{affect}} the & Verif*, Validat*, Test & Method \\
    & \textbf{\underline{verification}} and \textbf{\underline{validation}} & E2E, End-to-End & Approach \\  
    & \textbf{\underline{strategy}} (e.g., \textbf{\underline{E2E}} vs Modular approach)? & ADS, AV & Technique \\     
    \cline{1-2}\cline{4-4} \multirow{4}{*}{Q8} & How to \textcolor{gray}{\dashuline{argue}} on \textbf{\underline{completeness}} of & Autonomous Driving & Completeness\\
    & \textbf{\underline{verification}} and \textbf{\underline{validation}} & Automated Driving & Adequacy\\
    & for an \textbf{\underline{E2E}}-enabled \textbf{\underline{ADS}}? & Autonomous Vehicle & Coverage\\ 
    \hline
\end{tabular}
\label{tab:keywords}
\end{table*}

\begin{table*}[tbp]
\centering
\caption{Search results from each literature repository, together with the aggregated totals and the number of unique records after removing duplicates. The final column reports the number of studies included for each review question.}
\begin{tabular}{|C{0.03\textwidth}|C{0.08\textwidth}|C{0.08\textwidth}|C{0.08\textwidth}|C{0.08\textwidth}|C{0.14\textwidth}|C{0.08\textwidth}|}
    \hline 
    \textbf{\#}  & 
    \textbf{IEEE} & 
    \textbf{ACM} &
    \textbf{Scopus} &
    \textbf{arXiv} &
    \textbf{Total / Unique} &
    \textbf{Included} \\
    \hline
    Q7 & 354 & 316 & 388 & 450 & 1508 / 1230  & 16 \\
    \hline
    Q8 & 7 & 104 & 3 & 47 & 161 / 159  & 1 \\
    \hline
\end{tabular}
\label{tab:results}
\end{table*}

\subsection{Search Keywords}
\label{sec:results:papers:keywords}

We first derived search keywords from the two selected review questions. Question Q7 concerns how the transition in system architecture, from modular to E2E, influences testing approaches, whereas Q8 addresses how to substantiate the completeness of testing. A preliminary search used keywords closely reflecting these questions to assess the availability of relevant literature. However, this search yielded only a limited number of studies. After discussions with the case company, we agreed to broaden the keywords to increase the coverage of potentially relevant literature. The final keywords are presented in Table~\ref{tab:keywords}. For Q7, the keywords emphasize methods, approaches, and techniques for testing E2E ADS, while for Q8, they focus on concepts related to testing completeness, such as adequacy and coverage, of E2E ADS.

\subsection{Search Results}
\label{sec:results:papers:selected}

The literature search used the keywords in Table~\ref{tab:keywords}. For IEEE Xplore~\cite{ieee}, ACM Digital Library~\cite{acm}, and Scopus~\cite{scopus}, the results were restricted to peer-reviewed conference and journal publications in English published between 2016 and 2026. For arXiv~\cite{arxiv}, the search was restricted to pre-print papers uploaded during the same period. After removing duplicates, a total of 1,230 papers remained for Q7 and 159 for Q8. The selection process proceeded in five stages:
\begin{enumerate}
    \item \textbf{Title} screening.
    \item \textbf{Detailed relevance check} using abstracts, introductions, and conclusions.
    \item \textbf{Inclusion}: primary studies explicitly addressing the selected review questions.
    \item \textbf{Exclusion}: non-ADS domains, non-testing work, or studies not explicitly targeting E2E ADS.
    \item \textbf{Cross-validation} among the authors to ensure selection consistency.
\end{enumerate}
This yielded 16 papers for Q7 and 1 for Q8 (Table III). The complete list is available on Zenodo~\cite{song_2026_19627023}.

\begin{table*}[tbp]
\centering
\caption{Overview of selected papers for Q7: Approaches for testing E2E ADS. }
\begin{tabular}{|C{0.03\textwidth}|L{0.13\textwidth}|L{0.05\textwidth}|L{0.18\textwidth}|L{0.2\textwidth}|L{0.2\textwidth}|}
    \hline 
    \textbf{\#}  &
    \textbf{Authors} &
    \textbf{Year} &
    \textbf{Proposed intervention} &
    \textbf{Paper's Aim} &
    \textbf{Approach Used}\\
    \hline
    P1 & Yousefizadeh et al. & 2025 & CoCoMagic & Critical scenario generation & Constraint/search-based \\ \hline
    P2 & Lu et al. & 2024 & EpiTESTER & Critical scenario generation & Constraint/search-based \\ 
    \hline
    P3 & Zhong et al. & 2023 & AutoFuzz & Critical scenario generation & Constraint/search-based \\
    \hline
    P4 & Li et al. & 2025 & CCTest & Critical scenario generation & Constraint/search-based \\
    \hline
    P5 & Ji et al. & 2025 & SuS\&ApMCMC & Critical scenario generation & Constraint/search-based \\
    \hline
    P6 & Geng et al. & 2025 & Driving in Corner Case & Critical scenario generation & Constraint/search-based \\
    \hline
    P7 & Yan et al. & 2025 & OSG & Critical scenario generation & Constraint/search-based \\
    \hline
    P8 & Cheng et al. & 2025 & STCLocker & Critical scenario generation & Constraint/search-based \\
    \hline
    P9 & Haq et al. & 2023 & MARLOT & Critical scenario generation & Reinforcement-learning-based \\
    \hline
    P10 & Liang et al. & 2025 & MARL-OT & Critical scenario generation & Reinforcement-learning-based \\
    \hline
    P11 & Wu et al. & 2025 & MOEQT & Critical scenario generation & Reinforcement-learning-based \\
    \hline
    P12 & Yang et al. & 2023 & Suicidal Pedestrian & Critical scenario generation & Reinforcement-learning-based \\
    \hline
    P13 & Zheng et al. & 2025 & CADiffusion & Critical scenario generation & Generative-model-based \\
    \hline
    P14 & Zhou et al. & 2025 & SafeMVDrive & Critical scenario generation & Generative-model-based \\
    \hline
    P15 & Wang et al. & 2023 & Driving into Danger & Critical scenario generation & Adversarial perturbation \\
    \hline
    P16 & Zhang et al. & 2024 & UniAda & Critical scenario generation & Adversarial perturbation \\
    \hline
\end{tabular}
\label{tab:results:q7}
\end{table*}

\section{Analysis and Practitioner Assessment of Selected Literature}
\label{sec:results:assessments}

The technological rules extracted from the selected papers were assessed by practitioners for their relevance and applicability within their specific industrial context. Because 21 practitioners participated and papers were presented sequentially, feedback may vary, reflecting different perspectives and evolving assessments over time. We report feedback on each paper here, while Section~\ref{sec:discussion:findings} presents aggregated perspectives by review question and approach.

\subsection{Q7 -- Approaches for E2E ADS Testing}
\label{sec:results:assessments:q7}

All 16 papers included for Q7 focus primarily on generating \textbf{critical scenarios} for testing E2E ADS. Although they use terms such as \textit{safety-critical, adversarial, risky, and failure} scenarios, they address the same broad objective: producing particularly challenging scenarios for testing. We group the studies into four categories based on used methods (Table~\ref{tab:results:q7}):
\begin{enumerate}
    \item Constraint/search-based
    \item Reinforcement-learning-based
    \item Generative-model-based
    \item Adversarial perturbation-based
\end{enumerate} 
Hybrid studies are classified by their primary method. For each paper, we extracted a technological rule and presented it to practitioners, who assessed its usefulness and applicability in the case-company context.

\begin{table*}[tbp]
\centering
\caption{Q7: Critical scenario generation papers with Constraint/Search-based approaches. }
\begin{tabular}{|C{0.03\textwidth}|L{0.28\textwidth}|L{0.28\textwidth}|L{0.25\textwidth}|}
    \hline 
    \textbf{\#}  &
    \textbf{What does the approach take?} &
    \textbf{What does the approach manipulate?} &
    \textbf{What does the approach produce?} \\
    \hline
    P1 & Source-scenario space, perturbation space & General traffic scene and environment & Divergence-revealing scenario pair \\
    \hline
    P2 & Configurable scenario parameters & General traffic scene and environment	& Safety-critical driving scenarios \\
    \hline
    P3 & Grammar-based scenario specification & General traffic scene and environment & Violation-inducing driving scenarios \\
    \hline
    P4 & Configurable scenario parameters & Traffic interaction/vehicle behavior &	Safety-critical driving scenarios \\
    \hline
    P5 & Real car-following traffic data & Traffic interaction/vehicle behavior	& Rare failure-prone scenarios \\
    \hline
    P6 & Real traffic scene, conditioning prompt & Traffic interaction/vehicle behavior & Adversarial driving scene/images \\
    \hline
    P7 & Real traffic data, scenario specification & Traffic interaction/vehicle behavior & Traffic scenarios of varying risks \\
    \hline
    P8 &  Multi-AV traffic scenario & Traffic interaction/vehicle behavior & Deadlock-inducing multi-AV scenario \\
    \hline
\end{tabular}
\label{tab:csg:search-based}
\end{table*}

\subsubsection{Constraint/search-based}
\label{sec:results:assessments:q7:search}

Eight papers (Table~\ref{tab:csg:search-based}) use constraint/search-based approaches to explore the scenario space and generate critical scenarios in simulation.

\begin{tcolorbox}[
    enhanced,
    sharp corners,
    colback=black!5!white,
    frame hidden,
    colbacktitle=black,
    coltitle=white,
    fonttitle=\bfseries,
    title={Technological rule 1 [P1]}
]
  To detect realistic behavioral degradations between versions of an ADS, apply constrained co-evolutionary metamorphic differential testing with interpretable rule-based root cause analysis.
\end{tcolorbox}

\noindent \textbf{P1}~\cite{P1} presents CoCoMagic, a method that defines metamorphic rules to describe expected behavioral changes under scenario perturbations and compares two system versions to identify differential violations. It employs cooperative co-evolutionary search, where one population evolves source scenarios and another evolves perturbations; their combinations form test scenarios. The search prioritizes cases that maximize differences in rule violations, while constraints and guided initialization ensure realism. Practitioners offered initial feedback, noting that the approach aligns with and could support regression testing of evolving system versions.

\begin{tcolorbox}[
    enhanced,
    sharp corners,
    colback=black!5!white,
    frame hidden,
    colbacktitle=black,
    coltitle=white,
    fonttitle=\bfseries,
    title={Technological rule 2 [P2]}
]
  To efficiently discover safety-critical scenarios in ADS testing, apply an epigenetic genetic algorithm that uses an attention-based model to dynamically silence less important environmental parameters and focus the search on the most influential ones.
\end{tcolorbox}

\noindent \textbf{P2}~\cite{P2} presents EpiTESTER, a method for generating dangerous driving scenarios using an enhanced genetic algorithm (epiGA). It iteratively evolves scenarios through selection and recombination, incorporating a gene-silencing mechanism guided by a transformer-based attention model. Less important parameters are temporarily fixed, enabling the search to focus on critical ones and improving efficiency in identifying high-risk scenarios. Practitioners found the approach highly useful, highlighting its relevance for managing the expanding scenario database for E2E ADS. In particular, it supports the identification of critical and long-tail scenarios and aligns with ongoing efforts leveraging real-world data and scenario metadata.

\begin{tcolorbox}[
    enhanced,
    sharp corners,
    colback=black!5!white,
    frame hidden,
    colbacktitle=black,
    coltitle=white,
    fonttitle=\bfseries,
    title={Technological rule 3 [P3]}
]
  To efficiently find diverse traffic violations in ADS simulation testing, apply grammar-guided evolutionary fuzzing with neural-network-guided seed selection and mutation.
\end{tcolorbox}

\noindent \textbf{P3}~\cite{P3} presents AutoFuzz, a grammar-guided evolutionary fuzzer for ADS that generates semantically valid driving scenarios and searches for diverse traffic violations. It leverages simulator API grammars to ensure validity, evolutionary search to explore the scenario space, and an incrementally trained neural network to guide seed selection and mutation toward inputs likely to trigger unique violations. Practitioners found the approach promising but noted practical limitations. In particular, their current infrastructure does not yet support large-scale, arbitrary simulation or scenario manipulation, as many scenarios remain log-based and are not fully queryable or modifiable. While the approach is relevant, its adoption depends on further development of the engineering toolchain and the availability of a unified framework to evaluate and compare such methods.

\begin{tcolorbox}[
    enhanced,
    sharp corners,
    colback=black!5!white,
    frame hidden,
    colbacktitle=black,
    coltitle=white,
    fonttitle=\bfseries,
    title={Technological rule 4 [P4]}
]
  To reveal weaknesses in ADS in conflict situations, apply Critical-Configuration Testing that computes the hardest manageable scenarios from vehicle-dynamics constraints and evaluates the ADS in simulation.
\end{tcolorbox}

\noindent \textbf{P4}~\cite{P4} presents CCTest, an approach that constructs the hardest yet theoretically manageable scenarios, particularly in yield/merge conflicts and traffic-light intersections. It analytically derives critical configurations (e.g., vehicle positions and velocities) based on vehicle dynamics constraints such as braking distance and acceleration, and evaluates whether the ADS can safely handle these boundary cases. Practitioners considered the approach highly valuable and aligned with their current efforts. Such boundary scenarios effectively reveal system weaknesses or improvements, as performance differences are most evident near safety limits (e.g., cases where applying autonomous emergency braking may avoid a collision, while alternative actions would not).

\begin{tcolorbox}[
    enhanced,
    sharp corners,
    colback=black!5!white,
    frame hidden,
    colbacktitle=black,
    coltitle=white,
    fonttitle=\bfseries,
    title={Technological rule 5 [P5]}
]
  To evaluate the safety of black-box ADS in rare but realistic traffic situations, apply subset simulation to estimate failure probability and adaptive MCMC to generate increasingly dangerous samples efficiently.
\end{tcolorbox}

\noindent \textbf{P5}~\cite{P5} proposes an efficient approach to estimate the failure probability of a black-box ADS under realistic traffic conditions. It first learns a scenario distribution from real-world data, evaluates scenarios in simulation, and assigns risk scores. It then applies subset simulation (SuS) to decompose rare failure events into intermediate levels, and uses adaptive Markov Chain Monte Carlo (ApMCMC) to generate progressively rarer and more critical scenarios. This enables faster and more accurate estimation of failure probabilities. Practitioners noted that similar techniques (e.g., SuS and MCMC) are already used in their scenario-based, coverage-driven testing efforts. However, they highlighted the challenge of high-dimensional scenario spaces as a key limitation of such approaches. 

\begin{tcolorbox}[
    enhanced,
    sharp corners,
    colback=black!5!white,
    frame hidden,
    colbacktitle=black,
    coltitle=white,
    fonttitle=\bfseries,
    title={Technological rule 6 [P6]}
]
  To evaluate E2E ADS in realistic corner-case scenarios, apply closed-loop adversarial testing that generates adversarial traffic behaviors and renders them into real-world images with a diffusion-based generator.
\end{tcolorbox}

\noindent \textbf{P6}~\cite{P6} proposes a closed-loop adversarial evaluation platform for E2E driving, integrating adversarial traffic flow generation, a diffusion-based image generator, and the E2E model under test. Traffic flow models generate surrounding behaviors, which are translated into realistic driving images for the model. A two-episode strategy is used to create corner cases: first, the model is evaluated on nominal traffic; then, adversarial yet plausible trajectories are selected based on collision risk, realism, and smoothness, and re-evaluated using generated images. Practitioners expressed concerns about the complexity of this multi-stage, hybrid approach but showed interest in its generative component, particularly the extension to other sensor modalities (e.g., LiDAR point clouds) beyond images.

\begin{tcolorbox}[
    enhanced,
    sharp corners,
    colback=black!5!white,
    frame hidden,
    colbacktitle=black,
    coltitle=white,
    fonttitle=\bfseries,
    title={Technological rule 7 [P7]}
]
  To generate diverse driving scenarios with controllable risk levels for automated driving testing, apply a risk-regulated search that combines learned naturalness estimation, quantitative risk indication, and speciation-based heuristic exploration.
\end{tcolorbox}

\noindent \textbf{P7}~\cite{P7} proposes OSG, a framework for generating diverse driving scenarios with controllable risk levels for ADS testing. It employs a MAF (Masked Autoregressive Flow)-based Naturalness Estimator trained on real-world data to assess realism, and a Risk Indicator (e.g., time-to-collision, collisions) to quantify danger. These are integrated via a Risk Intensity Regulator in the objective function, enabling targeted generation of scenarios with varying risk levels. A speciation-based heuristic search further promotes diversity by exploring multiple regions of the scenario space in parallel. Practitioners considered the approach well-founded, particularly due to its integration of real-world data and multi-objective optimization to generate realistic and meaningful test scenarios.

\begin{tcolorbox}[
    enhanced,
    sharp corners,
    colback=black!5!white,
    frame hidden,
    colbacktitle=black,
    coltitle=white,
    fonttitle=\bfseries,
    title={Technological rule 8 [P8]}
]
  To efficiently find multi-vehicle deadlock scenarios in autonomous driving testing, apply feedback-guided search that detects wait-for cycles and mutates routes and trigger times based on spatial and temporal conflict.
\end{tcolorbox}

\noindent \textbf{P8}~\cite{P8} presents STCLocker, a feedback-guided search-based testing method for identifying multi-vehicle deadlock scenarios. It includes a Deadlock Oracle to detect wait-for cycles among vehicles, a Conflict Feedback mechanism to measure proximity to deadlock based on spatial and temporal interactions, and a Conflict-aware Scenario Generation process that iteratively mutates routes and timing before re-simulation. This loop progressively exposes weak cooperative behavior among multiple vehicles. Practitioners considered such scenarios less critical than collision-based cases and therefore not a current priority. However, they noted that the approach may become more relevant in large-scale, multi-vehicle environments.

\begin{table*}[tbp]
\centering
\caption{Q7: Critical scenario generation papers with Reinforcement-learning-based approaches. }
\begin{tabular}{|C{0.03\textwidth}|L{0.28\textwidth}|L{0.28\textwidth}|L{0.25\textwidth}|}
    \hline 
    \textbf{\#}  &
    \textbf{What does the approach take?} &
    \textbf{What does the approach manipulate?} &
    \textbf{What does the approach produce?} \\
    \hline
    P9 & Simulated driving scene, requirements & General traffic scene and environment & Violation-inducing driving scenario \\
    \hline
    P10 & Simulated driving scene & Traffic interaction/vehicle behavior & Safety-critical driving scenario \\
    \hline
    P11 & Simulated driving scene, requirements & Traffic interaction/vehicle behavior & Violation-inducing driving scenario \\
    \hline
    P12 & Simulated driving scene & Pedestrian behavior & Pedestrian safety-critical scenario \\
    \hline
\end{tabular}
\label{tab:csg:rl-based}
\end{table*}

\subsubsection{Reinforcement learning-based} 

Four studies (Table~\ref{tab:csg:rl-based}) employ reinforcement learning to optimize sequences of actions or environmental changes that generate critical scenarios.
\label{sec:results:assessments:q7:rl}

\begin{tcolorbox}[
    enhanced,
    sharp corners,
    colback=black!5!white,
    frame hidden,
    colbacktitle=black,
    coltitle=white,
    fonttitle=\bfseries,
    title={Technological rule 9 [P9]}
]
  To efficiently find failures in online testing of DNN-enabled systems, apply many-objective reinforcement learning that targets the uncovered requirement closest to violation.
\end{tcolorbox}

\noindent \textbf{P9}~\cite{P9} proposes MORLOT, an online testing approach for DNN-based systems that identifies failure scenarios by incrementally modifying the environment during simulation. It combines many-objective search to prioritize the requirement closest to violation with reinforcement learning to select environment changes that drive the system toward violating that requirement. Practitioners found the generated scenarios difficult to interpret and noted that focusing solely on requirement violations (e.g., speeding) may produce unrealistic scenarios. They emphasized the need for additional constraints to ensure the generated scenarios remain meaningful and plausible.

\begin{tcolorbox}[
    enhanced,
    sharp corners,
    colback=black!5!white,
    frame hidden,
    colbacktitle=black,
    coltitle=white,
    fonttitle=\bfseries,
    title={Technological rule 10 [P10]}
]
  To efficiently detect safety violations in ADS, apply multi-agent reinforcement learning to coordinate surrounding vehicles and trigger rule-based online fuzzing when risky interaction patterns appear.
\end{tcolorbox}

\noindent \textbf{P10}~\cite{P10} presents MARL-OT, an online testing method that identifies safety violations through coordinated multi-vehicle behavior. It combines a pre-trained multi-agent reinforcement learning model to guide surrounding vehicles into critical configurations with a rule-based fuzzer that executes realistic maneuvers (e.g., braking, lane changes) once risky situations emerge. Practitioners found the approach highly relevant, noting its alignment with ongoing efforts to train RL-based planners and move from static to reactive, closed-loop testing. It supports evaluating how traffic responds to ADS behavior and enables testing of both naturalistic and safety-critical multi-vehicle scenarios, depending on the objectives.

\begin{tcolorbox}[
    enhanced,
    sharp corners,
    colback=black!5!white,
    frame hidden,
    colbacktitle=black,
    coltitle=white,
    fonttitle=\bfseries,
    title={Technological rule 11 [P11]}
]
  To generate critical scenarios that violate multiple interdependent requirements in ADS testing, apply multi-objective reinforcement learning with dynamically balanced objective weights.
\end{tcolorbox}

\noindent \textbf{P11}~\cite{P11} presents MOEQT, a testing method that uses multi-objective reinforcement learning to generate scenarios violating multiple interdependent requirements. It iteratively observes system and environment states, applies actions to modify the environment, evaluates each requirement via dedicated objective functions, and learns using Envelope Q-learning to balance objectives and identify critical scenarios. Similar to P9, practitioners questioned the choice of requirements and the realism of the generated scenarios, leading to uncertainty regarding their practical usefulness for testing.

\begin{tcolorbox}[
    enhanced,
    sharp corners,
    colback=black!5!white,
    frame hidden,
    colbacktitle=black,
    coltitle=white,
    fonttitle=\bfseries,
    title={Technological rule 12 [P12]}
]
  To generate pedestrian-related critical scenarios for ADS testing, apply reinforcement learning to train an adversarial pedestrian with collision-focused rewards.
\end{tcolorbox}

\noindent \textbf{P12}~\cite{P12} generates pedestrian-related safety-critical scenarios by training a reinforcement learning pedestrian agent to induce collisions with the autonomous vehicle. The agent observes relative position and motion, selects actions (e.g., direction and speed), and is trained using PPO with collision-focused rewards to produce diverse critical scenarios. Practitioners considered the approach useful for enriching scenario sets, although it is not currently in use; it remains a potential direction for future exploration.

\subsubsection{Generative model-based} 
\label{sec:results:assessments:q7:generative}

Two studies (in Table~\ref{tab:csg:genreative-based}) use recent generative models to generate desired critical scenarios.

\begin{table*}[tbp]
\centering
\caption{Q7: Critical scenario generation papers with Generative-model-based approaches. }
\begin{tabular}{|C{0.03\textwidth}|L{0.28\textwidth}|L{0.28\textwidth}|L{0.25\textwidth}|}
    \hline 
    \textbf{\#}  &
    \textbf{What does the approach take?} &
    \textbf{What does the approach manipulate?} &
    \textbf{What does the approach produce?} \\
    \hline
    P13 & Driving image, control conditions & Lane/lane-marking & Lane-focused adversarial driving image \\
    \hline
    P14 & Multi-view driving scene, conditioning prompt & Traffic interaction/vehicle behavior & Multi-view safety-critical driving video \\
    \hline
\end{tabular}
\label{tab:csg:genreative-based}
\end{table*}

\begin{table*}[tbp]
\centering
\caption{Q7: Critical scenario generation papers with Adversarial perturbation approaches. }
\begin{tabular}{|C{0.03\textwidth}|L{0.28\textwidth}|L{0.28\textwidth}|L{0.25\textwidth}|}
    \hline 
    \textbf{\#}  &
    \textbf{What does the approach take?} &
    \textbf{What does the approach manipulate?} &
    \textbf{What does the approach produce?} \\
    \hline
    P15 & Driving image with pedestrian & Pedestrian clothing & Pedestrian safety-critical scenarios \\
    \hline
    P16 & Driving video/image sequence & Pixel/image perturbations & Adversarial images/perturbed images \\
    \hline
\end{tabular}
\label{tab:csg:perturbation-based}
\end{table*}

\begin{tcolorbox}[
    enhanced,
    sharp corners,
    colback=black!5!white,
    frame hidden,
    colbacktitle=black,
    coltitle=white,
    fonttitle=\bfseries,
    title={Technological rule 13 [P13]}
]
  To generate realistic and controllable adversarial images for testing lane detection in ADS, apply diffusion-based image editing guided by BLIP semantic prompts, spatial masks, and ControlNet edge constraints.
\end{tcolorbox}

\noindent \textbf{P13}~\cite{P13} presents CADiffusion, a diffusion-based framework for generating adversarial road images under semantic, spatial, and geometric constraints. It uses BLIP and text prompts to define perturbations, a mask to localize edits, and ControlNet to preserve road geometry, producing realistic images that maintain scene coherence while disrupting lane detection. Practitioners considered the approach novel and valuable, particularly for analyzing perception failures and downstream effects on trajectory planning in E2E ADS. Its controllability enables targeted perturbations (e.g., vision-only), making it useful for isolating sensor-specific effects in multi-sensor fusion and supporting systematic evaluation of system behavior.

\begin{tcolorbox}[
    enhanced,
    sharp corners,
    colback=black!5!white,
    frame hidden,
    colbacktitle=black,
    coltitle=white,
    fonttitle=\bfseries,
    title={Technological rule 14 [P14]}
]
  To generate multi-view safety-critical driving videos for testing E2E ADS, apply VLM-guided adversarial vehicle selection with two-stage diffusion-based trajectory generation and multi-view video synthesis.
\end{tcolorbox}

\noindent \textbf{P14}~\cite{P14} presents SafeMVDrive, an approach for generating realistic multi-view safety-critical driving videos for testing E2E ADS. It uses a GRPO-finetuned vision-language model to identify the most dangerous vehicle, followed by a two-stage diffusion-based trajectory generator trained on real-world data to produce collision and near-collision scenarios. These trajectories are then converted into high-quality multi-view videos for stress testing. Practitioners found the approach promising but challenging to adopt due to its complexity and reliance on multiple components and infrastructures. While it aligns with ongoing efforts (e.g., using VLMs for scenario extraction), integrating such a multi-stage pipeline into existing toolchains and testing strategies requires significant engineering effort, making it a potential direction for future adoption.

\subsubsection{Adversarial perturbation-based} 
\label{sec:results:assessments:q7:perturbation}

Two studies, shown in Table~\ref{tab:csg:perturbation-based}, employ adversarial perturbation-based approaches to transform existing scenarios into critical ones.

\begin{tcolorbox}[
    enhanced,
    sharp corners,
    colback=black!5!white,
    frame hidden,
    colbacktitle=black,
    coltitle=white,
    fonttitle=\bfseries,
    title={Technological rule 15 [P15]}
]
  To expose unsafe behavior in E2E ADS when pedestrians are present, apply calibrated adversarial patch optimization on pedestrian clothing in simulation.
\end{tcolorbox}

\noindent \textbf{P15}~\cite{P15} proposes an adversarial T-shirt patch attack on an E2E ADS. It uses CARLA simulation and UE4-based texture editing to determine patch placement, deformation, and color calibration for realism. The patch is then optimized using a loss on waypoints and control signals to induce unsafe behavior when applied to a pedestrian’s clothing in simulation. Practitioners found this approach highly relevant, as similar issues have been observed in practice. For example, clothing colors can affect perception, causing pedestrians to be partially or entirely missed (e.g., blending with the background). They noted that such edge cases are consistent with their efforts to analyse the impact of color and appearance on perception, particularly when tuning environmental and object properties.

\begin{tcolorbox}[
    enhanced,
    sharp corners,
    colback=black!5!white,
    frame hidden,
    colbacktitle=black,
    coltitle=white,
    fonttitle=\bfseries,
    title={Technological rule 16 [P16]}
]
To test E2E ADS for steering and acceleration failures, apply white-box gradient-based multi-objective optimization to learn a universal pixel-level perturbation with adaptive objective weighting.
\end{tcolorbox}

\noindent \textbf{P16} (UniAda~\cite{P16}) is a white-box, gradient-based adversarial testing method for E2E ADS. It optimizes a universal pixel-level perturbation across multiple video frames to mislead both steering and acceleration, using iterative multi-objective optimization with adaptive weighting to balance attack goals. Practitioners questioned the realism of such pixel-level perturbations, noting that while effective in simulation, it is unclear whether the generated variations are physically meaningful or representative of real-world conditions as what the sensors (e.g., cameras) really capture. This raises concerns about their relevance for practical testing scenarios.

\subsection{Q8 -- Completeness of E2E ADS Testing}
\label{sec:results:assessments:q8}

Only one study was included in the analysis for Q8, as presented in Table~\ref{tab:results:q8}. In general, the study addresses the completeness of testing for E2E ADS through a scenario-based, data-driven approach aimed at maximizing test coverage.

\begin{table*}[tbp]
\centering
\caption{Overview of selected papers for Q8: Completeness of testing E2E ADS. }
\begin{tabular}{|C{0.03\textwidth}|L{0.13\textwidth}|L{0.05\textwidth}|L{0.18\textwidth}|L{0.2\textwidth}|L{0.2\textwidth}|}
    \hline 
    \textbf{\#}  &
    \textbf{Authors} &
    \textbf{Year} &
    \textbf{Proposed intervention} &
    \textbf{Paper's aim} &
    \textbf{Approach used}\\
    \hline
    P17 & Pathrudkar et al. & 2023 & SAFR-AV & Maximize test coverage & Scenario-based, Data-driven \\
    \hline
\end{tabular}
\label{tab:results:q8}
\end{table*}

\begin{table*}[tbp]
\centering
\caption{Q8: Papers with Scenario-based and data-driven approaches. }
\begin{tabular}{|C{0.03\textwidth}|L{0.28\textwidth}|L{0.28\textwidth}|L{0.25\textwidth}|}
    \hline 
    \textbf{\#}  &
    \textbf{What does the approach take?} &
    \textbf{What does the approach learn?} &
    \textbf{What does the approach produce?} \\
    \hline
    P17 & Real-world driving data & Joint distributions of scenario parameters & Plausible scenario variations \\
    \hline
\end{tabular}
\label{tab:completness}
\end{table*}

\begin{tcolorbox}[
    enhanced,
    sharp corners,
    colback=black!5!white,
    frame hidden,
    colbacktitle=black,
    coltitle=white,
    fonttitle=\bfseries,
    title={Technological rule 17 [P17]}
]
  To achieve more systematic test coverage for ADS, apply behavioral-competency-based scenario selection from real-world data and joint parameter distribution learning to generate plausible scenario variations.
\end{tcolorbox}

\noindent \textbf{P17}~\cite{P17} proposes a scenario-based, data-driven approach to support arguments for testing coverage and completeness in ADS. It models behavioral competencies (e.g., lane changes), retrieves corresponding real-world scenarios, and learns joint parameter distributions to generate plausible variations. While it does not ensure absolute completeness, it enables more systematic coverage of the scenario and parameter space. Practitioners considered the approach highly relevant, as it aligns with ongoing coverage-driven testing using real-world data. However, key challenges remain, including defining coverage targets and assessing achievable coverage. Scenario complexity leads to high-dimensional spaces, making comprehensive coverage difficult, while statistical modeling requires capturing complex correlations and nonlinearities. This highlights the need for robust models and metrics to quantify coverage.

\section{Discussion}
\label{sec:discussion}


\subsection{Findings for Review Questions}
\label{sec:discussion:findings}

\subsubsection{Approaches for Testing}
\label{sec:discussion:findings:q7}
We identified 16 studies addressing Q7, all of which focus on generating critical scenarios for testing E2E ADS using different approaches. 

\underline{Overall relevance.} Practitioners’ assessments indicate that the majority of these approaches are relevant and useful, with some already reflecting practices adopted in industry. Across categories, proposed approaches support the identification of \textit{critical, long-tail, unsafe, and boundary scenarios} that expose system weaknesses, thereby enriching scenario sets for testing. This aligns well with ongoing industrial efforts that combine real-world data with synthetic scenario generation. 

\underline{Realism and rationality.} A key theme emerging from practitioner feedback is the importance of \textit{realism and rationality of generated scenarios}. Approaches that incorporate real-world data to learn scenario distributions, enforce constraints or control conditions during generation, or model naturalistic multi-agent interactions were generally preferred. This includes search-based, reinforcement learning-based, and generative approaches that produce plausible and meaningful scenarios, as well as perturbation methods that apply realistic modifications. Such characteristics are considered essential for ensuring that generated scenarios are representative of real-world driving conditions and thus valuable for testing.

\underline{Adoption barriers.} However, several challenges limit practical adoption of these interventions. Practitioners consistently questioned the \textit{complexity and realism of the generated scenarios}, noting that even simple cases require high-dimensional parameter spaces. It remains unclear which dimensions are considered in existing studies, how challenges such as the curse of dimensionality are addressed (e.g., in search-based approaches), how
simulation quality is assured, and whether the resulting scenarios faithfully represent realistic driving conditions. Further, many approaches involve \textit{complex, multi-stage pipelines or multiple tools}, making them difficult to integrate into existing toolchains and testing workflows. In addition, current infrastructure may also lack the capability to support large-scale, flexible simulation, scenario manipulation, and variation generation, hindering immediate replication of these approaches in their specific industrial contexts.

\underline{Comparative evaluation.} Despite these constraints,
practitioners considered several approaches promising and
potentially useful. They also emphasized the need for a \textit{unified framework} to systematically evaluate and compare different methods in order to better understand their relative strengths, limitations, and practical value, highlighting an important need and a direction for future research.

\subsubsection{Completeness of Testing}
\label{sec:discussion:findings:q8}

Although only one study was identified for Q8, the proposed approach was considered highly relevant by practitioners, as it aligns with ongoing efforts in scenario- and data-driven, coverage-based testing aimed at maximizing test coverage. Four fundamental questions remain unresolved:

\begin{itemize}
    \item How should appropriate coverage targets be defined?
    \item What level of coverage is realistically achievable?
    \item How can coverage be quantified in a meaningful way?
    \item How do such measures relate to testing completeness and safety assurance?
\end{itemize}

While the study provides a practical foundation and useful direction for future work, these open challenges, together with the limited number of identified studies, highlight the difficulty of this area despite its importance, and underscore the critical need for further research.

\subsection{Answers to Research Questions}
\label{sec:discussion:rqs}

\underline{RQ1 -- Key challenges.} We closely collaborated with the case company to identify practical challenges and their specific contexts. Practitioners proposed 12 open questions, primarily concerning \textit{testing completeness and testing of E2E ADS}, with additional questions on data quality, safety argumentation, and simulation realism. The two highest-ranked questions concern testing approaches and testing completeness for E2E ADS, indicating their urgency and importance from an industrial perspective. This also aligns with the growing adoption of E2E architectures in recent research and practice~\cite{10614862}.

\underline{RQ2 -- Relevant research.} We selected 17 papers that explicitly address the two review questions and extracted their technological rules. 16 studies focus on generating \textit{critical scenarios} for testing, while one proposes a scenario-based, data-driven approach to support arguments for testing completeness.

\underline{RQ3 -- Applicability and adoption.} We presented these papers and their corresponding rules to practitioners, who assessed their relevance and applicability in their industrial context. Most approaches were considered \textit{promising and relevant, with some already reflected in ongoing practices}, while others require further investigation, particularly regarding toolchain integration, scenario realism, and practical applicability.

\begin{tcolorbox}[
    enhanced,
    colback=black!5,
    frame hidden,
    borderline west={3pt}{0pt}{black},
    sharp corners,
    boxsep=0pt,
    left=10pt,
    right=10pt,
    top=8pt,
    bottom=8pt
]
\textbf{\textit{Overall Insight:}} Overall, these findings indicate that current research can provide value to practitioners while also pointing to the need for further work to bridge research and industrial adoption and for closer collaboration between academia and industry.
\end{tcolorbox}

\subsection{Reflections on Lessons Learned}
\label{sec:discussion:lessons}

Overall, we found the \textit{interactive rapid review} to be an effective method for gaining close insights into industrial practice and facilitating the transfer of research into industry. Four methodological lessons stand out from this collaborative experience:

\begin{itemize}
    \item \textbf{Enable rapid, bidirectional knowledge exchange.} The review enabled the identification of \textit{open challenges} faced by practitioners, revealing practical problems and highlighting important research directions in ADS testing. It also provided empirical insights into the perceived importance and urgency of these challenges, indicating what matters most to practitioners and where further research is needed. Beyond identifying challenges, the review \textit{offered timely input to industry} by synthesizing existing research and providing an overview of available approaches, thereby supporting informed and rapid decision-making. In return, practitioners contributed valuable \textit{feedback on the relevance, applicability, and potential impact} of these approaches, both in the short and long term. Such exchange is essential for evidence-based and context-aware research, helping to distinguish between practically relevant and less applicable contributions.
    \item \textbf{Sustain practitioner involvement.} Applying an interactive rapid review is inherently challenging, as it requires \textit{sustained practitioner involvement across multiple stages}, including sharing experiences, practical problems, contextual knowledge, and current practices, all of which are often difficult to obtain~\cite{song2023industry}. Strong industry–academia collaboration is therefore critical to the success of such studies~\cite{rico2024experiences}. In this work, 21 practitioners across diverse roles participated, including function owners, system architects, technical leads, engineers, designers, and testers, enhancing the breadth and value of the findings.
    \item \textbf{Use an explicit collaboration protocol.} A \textit{well-defined collaboration protocol}~\cite{rico2024experiences} is essential for coordinating activities and sustaining effective collaboration, and should include scheduling meetings in advance, clearly defining agendas and responsibilities, and fostering timely communication and active discussion.
    \item \textbf{Make participation valuable and appoint a champion.} It is equally important that practitioners understand the \textit{benefits of participation}, particularly how the outcomes relate to their work and what they can gain from the study.  Finally, having a managerial contact or committed champion within the case company who can coordinate activities and engage stakeholders contributes significantly to the success of the study and the quality of its outcomes. Together, these practices establish a strong foundation for continued collaboration, a shared understanding of problems and contexts, and further joint efforts between academia and industry.
\end{itemize}

\subsection{Threats to Validity}
\label{sec:discussion:ttv}

As an interactive rapid review conducted with a case company, \textit{internal validity} is of primary importance to this study.

\underline{Internal validity.} Internal validity concerns the credibility of findings within the study~\cite{verdecchia2023threats, lago2024threats}. A key threat lies in involving appropriate stakeholders who can provide relevant challenges, contextual knowledge, and accurate assessments of relevance. With the support of a committed champion within the case company, we were able to engage a substantial group of knowledgeable stakeholders. In particular, during the assessment phase, 21 practitioners participated, representing a wide range of roles and substantial experience in ADS testing. This broad and informed participation strengthens the internal validity of the findings. Another limitation concerns Q8, for which only one relevant study was identified. While this limits the available evidence for that question, it also reflects the limited research currently addressing this important challenge and highlights the need for further investigation.

\underline{Construct validity.} Threats to construct validity are considered limited because the study adopts a broad perspective on ADS testing rather than focusing on specific techniques or concepts. All participating stakeholders are actively involved in and experienced with ADS testing, supporting the appropriateness of the constructs examined. 

\underline{External validity.} Although external validity is not the primary focus of this study, the case company is a well-established automotive organization with extensive experience in ADS development and testing. Its active collaboration with other companies and participation in large-scale research initiatives involving diverse stakeholders suggest that the findings may be transferable to similar industrial contexts.

\section{Conclusion}
\label{sec:conclusion}

With the advancement and deployment of ADS in real traffic, testing has become both essential and inherently challenging due to open operational environments and the lack of universally established processes, approaches, and metrics. Through an interactive rapid review involving 21 practitioners from a world-leading automotive company, we identified practical challenges in ADS testing, reviewed relevant literature on testing E2E ADS, and assessed their relevance and applicability in an industrial context. 

Our findings indicate that, while most approaches reported in academic research are considered useful, several require further investigation before practical adoption. We conclude that systematic and effective scenario-based testing methods for E2E ADS are needed, and that demonstrating testing completeness remains a critical yet unresolved challenge. Finally, the study contributes to both academia and industry by enabling the exchange of insights and knowledge, supporting the resolution of practical testing challenges, and promoting greater realism and industrial relevance in future research.

\section*{Ethical Considerations}
The study was conducted in collaboration with the case company, which agreed to participate and invited relevant practitioners to take part. Practitioners were informed about the study and participated voluntarily. Their responses were anonymized and reported without identifying individuals.

\section*{Data Availability}
The data that support the findings of this study is available at Zenodo at \url{https://doi.org/10.5281/zenodo.19627023}.

\bibliographystyle{IEEEtran}
\bibliography{ref}

\end{document}